\documentclass[galaxies,review,accept,pdftex,oneauthor]{Definitions/mdpi}
\usepackage{upgreek}

\firstpage{1}
\pubvolume{13}
\issuenum{2}
\articlenumber{25}
\pubyear{2025}
\copyrightyear{2025}
\externaleditor{Dominic Bowman}
\datereceived{14 February 2025}
\daterevised{6 March 2025} 
\dateaccepted{11 March 2025}
\datepublished{20 March 2025}
\hreflink{https://doi.org/10.3390/galaxies\linebreak13020025} 

\Title{Red Supergiants---The Other Side of the H-R Diagram}

\TitleCitation{Red Supergiants---The Other Side of the H-R Diagram}

\Author{{Roberta} 
{M.} 
Humphreys \href{https://orcid.org/0000-0003-1720-9807}{\orcidicon}}

\AuthorNames{Roberta M. Humphreys}

\AuthorCitation{Humphreys, R.M.}

\address[1]{%
{Minnesota Institute for Astrophysics, University of
Minnesota, 116 Church St. SE, Minneapolis, MN 55455, USA}
; roberta@umn.edu}

\abstract{Red supergiants are the largest stars known with some of the highest mass loss rates observed. They are the final stage in the evolution of the majority of massive stars. The unexpected discovery of high mass loss episodes in many red supergiants have posed questions about the role of mass loss on their final stages. The papers in this volume are timely reviews of our current understanding of this often surprising population of massive stars.  This introductory paper is a brief summary of their observed properties and a historical perspective on some of the current problems on mass loss, their circumstellar environments, and their evolutionary state.}


\begin{document}


\section{Introduction}

For decades, most astronomers have considered red supergiants to be relatively well understood. They were post main-sequence massive stars in He-burning that would end their brief lives as
supernovae. But, along the way, the red supergiants have taught us about circumstellar dust and gas, stellar instabilities, mass loss, and its consequences for stellar evolution. In recent years, the instabilities and high mass loss episodes in the red supergiants and other massive stars have posed more uncertainties about their final stages and their terminal state.  The review
papers in this Special Issue focus on this often surprising population of massive stars. Emily Levesque's  book on red supergiants \citep{EL} is an excellent introduction to their observed properties, structure, and evolution. The topical reviews included here are a timely supplement covering the most current
questions about the red supergiants. The reviews cover the populations of red supergiants in different galaxies, mass loss and their circumstellar ejecta, dust and gas, special individual stars like Betelguese and VY CMa, post-red supergiant evolution, and as supernova progenitors.


In this introductory chapter, I give a brief overview of the red supergiants and
a historical perspective on the topics in this issue.

\section{Red Supergiants on the H-R Diagram\label{sec2}}

\begin{figure}[H]
\begin{adjustwidth}{-\extralength}{0cm}
\centering 

\includegraphics[width=8.0cm]{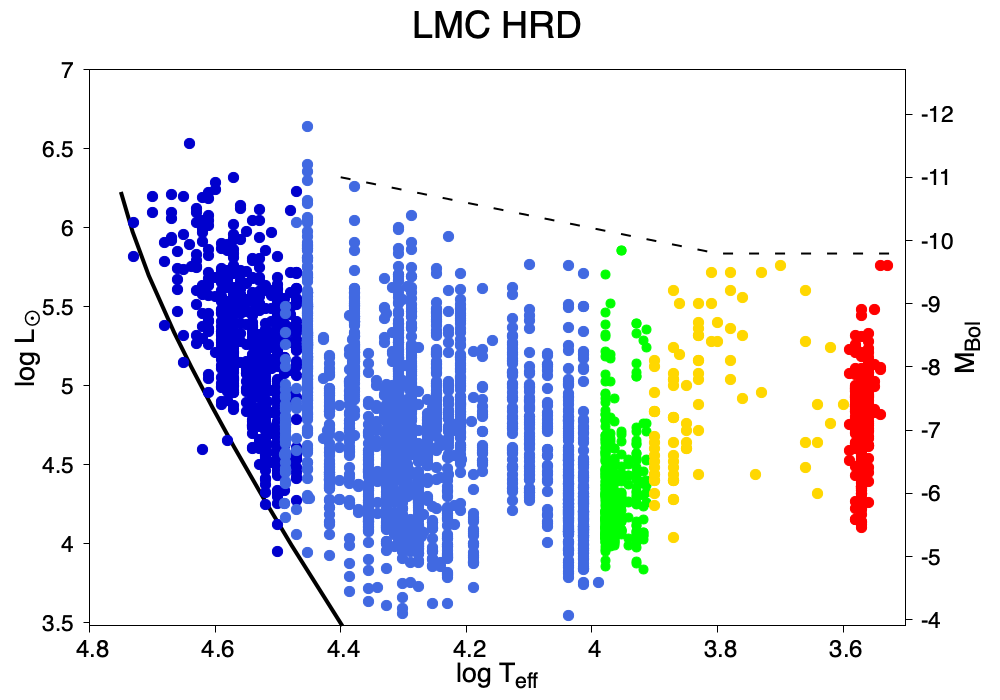}
%
\includegraphics[width=8cm]{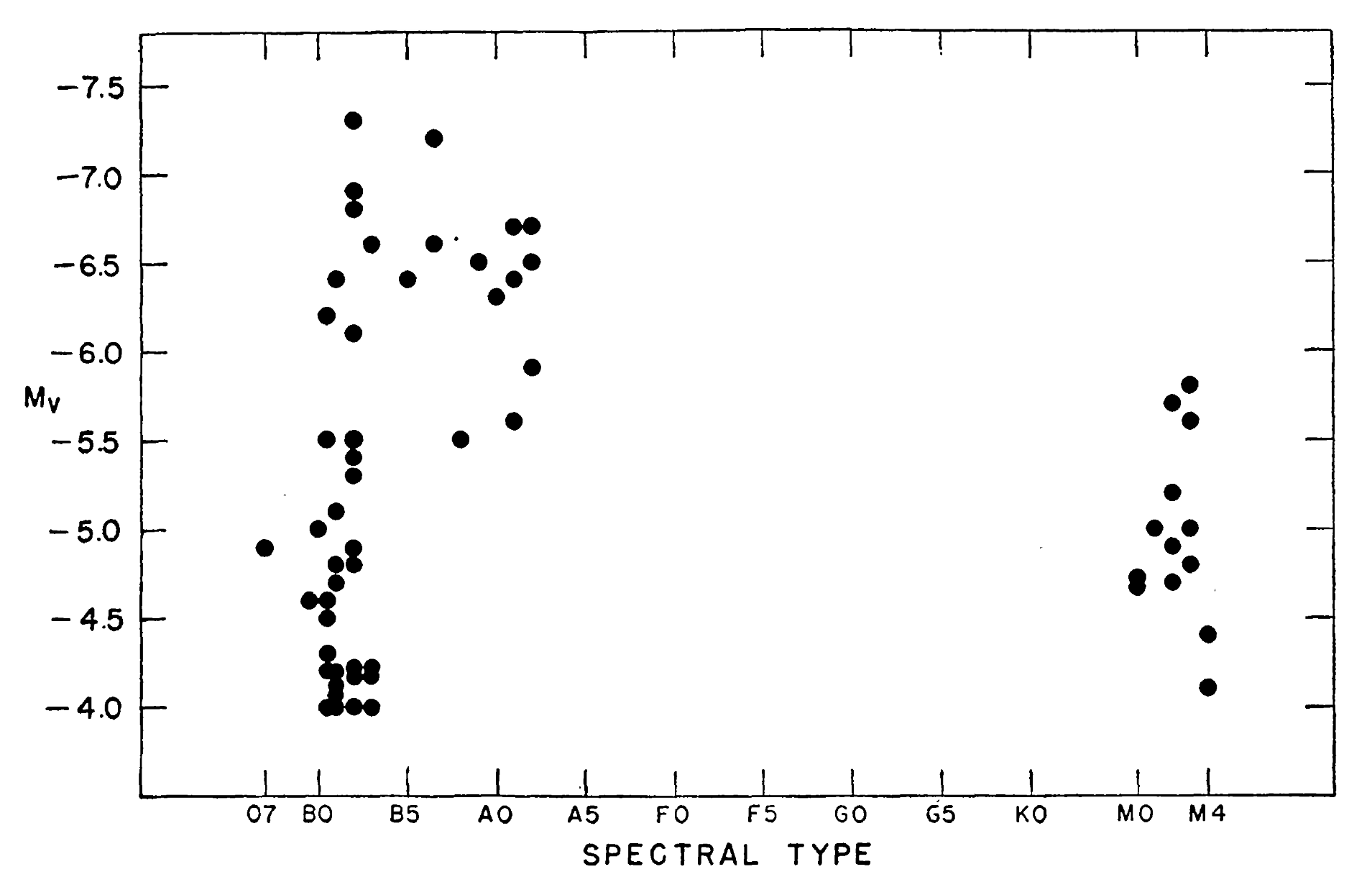}
\end{adjustwidth}
\caption{{(\textbf{Left})}
: A recent HRD for the luminous stars in the LMC with spectral types \citep{JM}. The dashed line is the Humphreys--Davidson (HD) limit. (\textbf{Right}): An early HRD for the Perseus cluster showing the blue and red supergiants \citep{WPB}.\label{fig1}}
\end{figure}

The  Hertzsprung--Russell Diagram (H-R Diagram, or HRD) for massive stars, whether in the Milky Way or other galaxies, is dominated by the large number of `hot'
stars on the `blue' side of the HRD; stars near the H-burning main sequence with spectral types O and B. On the `other side', there is a relatively narrow band with many fewer stars at significantly lower temperatures, the red supergiants
or M-type supergiants. Figure~\ref{fig1} shows a current H-R Diagram for the Large Magellanic Cloud (LMC) for luminous stars with spectral types~\citep{JM}, illustrating the relative populations of the two sides of the HRD. For comparison, an early H-R Diagram for h and $\chi$ Perseus, the Perseus association,
known for its large number of red supergiants, is also shown, illustrating the same effect but on a smaller scale. Of course, this dichotomy is well-understood due to the relative lifetimes of massive stars in H-burning and the H-shell burning stages as they evolve
across the HRD to the red supergiant stage for the onset of the shorter He-burning stage.  Massive stars are usually defined as those with initial masses above about 9 M$_{\odot}$ with sufficient mass in their cores for He-fusion.

But, not all massive stars become red supergiants.~\citet{HD79} recognized an
empirical upper luminosity boundary in the observed H-R Diagrams for the massive star populations in the Milky Way and LMC with an envelope of declining
luminosity with decreasing temperature for the hotter stars and a relatively
tight upper limit to the luminosities of the cooler evolved stars at about M$_{Bol}$ = $-$9.7, Log L/L$_{\odot}$ = 5.7. This flat upper limit was defined by the most luminous cool and intermediate temperature supergiants in both galaxies, the red and yellow hypergiants. The lack of these evolved stars above this luminosity demonstrated that massive stars, above about 40 M$_{\odot}$, based on evolutionary models, did not evolve across
the H-R Diagram to become red supergiants due to post-main sequence instabilities and high mass loss events. The temperature dependence of the upper boundary for the hotter stars supports a mass dependent instability.


The red and yellow hypergiants were critical to the recognition of this upper
limit in the H-R Diagram. Stars above 40 M$_{\odot}$ will not become red supergiants, but {\it {the majority of massive stars with initial masses from 9 to 40 M$_{\odot}$, will enter the red supergiant stage.} 
}


\section{So What Is a Red Supergiant?}

\subsection{Physical Properties}

Observationally, red supergiants are identified both by their color and by
their spectra with strong molecular bands primarily of TiO in the classical blue-visual and red spectral regions. But, they share these properties with the red giants and AGB stars. The supergiants are distinguished from these other groups of cool or red stars by their high luminosities and large size. The direct measurement of their luminosities depends on distance and their angular sizes were measured directly by interferometry for only a few stars.


The range of their observed and derived parameters discussed here are summarized in Table~\ref{tab1}.


The red supergiants are most readily separated from the other classes of red stars spectroscopically. The earliest classification of stellar spectra by Angelo
Secchi  included five groups. Group III was equivalent with the spectral type M  stars on the Harvard system \citep{McCarthy}. On the Harvard classification system \citep{HA},  the type  M stars with TiO bands include supergiants and giants, and based on the strength of the TiO bands, the spectral types ranged from M0 to M4.  Because luminosity differences were suspected, Morgan and Keenan \citep{MK} recognized and defined luminosity-based spectral criteria across the full range of stellar spectral types.  These criteria allowed the separation into luminosity classes, I--V, and the
spectroscopic separation of the M supergiants (I) from M giants (III) and M dwarfs
or main sequence stars (V).
\begin{table}[H]
\caption{{Range} 
of observed and derived parameters.\label{tab1}}
\centering
\newcolumntype{C}{>{\centering\arraybackslash}X}
\begin{tabularx}{\textwidth}{cCC}

\toprule
\textbf{Parameter}   & \textbf{Range} &  \textbf{Comment}\\
\midrule
Sp Type     & K5--M4.5   &  \\
T$_\mathrm{eff}$   & 3900--3400 K &  \\
Luminosity  &  M$_\mathrm{V}$ $\approx$ $-$4 to $-$8 (mag) &    \\
&  M$_\mathrm{Bol}$ $-$5.5 to $-$9.7 (mag)  &     \\
&  Log L/L$_{\odot}$ 4.5 to 5.7  &     \\
Radius      &  $\approx$1.5 to 7 AU  &    from CHARA interferometry  \\
&  300 to 1500 R$_{\odot}$ &     \\
Mass        &  9 to $\approx$40 M$_{\odot}$ & based on evolutionary tracks \\
\bottomrule
\end{tabularx}
\end{table}


In the Milky Way and other nearby galaxies, the observed parameters are primarily
spectra, and magnitudes and colors, often for a range of wavelengths from the visual to the near and mid-infrared.  Spectral type and color are both indicators of temperature. When the distance is known, the absolute visual magnitude, M$_\mathrm{V}$, corrected for interstellar extinction, versus the spectral type or intrinsic color is used in HRDs, especially in the early ones.  But, for many astrophysical applications, comparing stellar populations, with different masses and temperatures, the
total luminosity vs the effective temperature is preferred.
The two HRDs in Figure~\ref{fig1} are examples. For h and $\chi$ Persei, the axes are
the spectral type and the M$_\mathrm{V}$ derived from the observed visual magnitude,
corrected for interstellar extinction, and the distance to the double cluster.  While for the LMC, log L/L$_{\odot}$ or M$_\mathrm{Bol}$ depends on the calibrations of the effective temperature with spectral type and the bolometric corrections to M$_\mathrm{V}$, and of course the adopted distance to the LMC.


The effective temperatures (T$_\mathrm{eff}$) corresponding to the observed spectral
types can be calibrated using stellar atmosphere models. Various T$_\mathrm{eff}$ scales
for
stars of all spectral types have existed for some time.   See~\citet{Flower}
for an example and an update~\citep{Flower2}. A commonly adopted
temperature calibration \citep{Levesque1,Levesque2} for the M supergiants is based on the MARCS models which include corrections for the opacity from TiO and other \mbox{molecules \citep{Plez,Gus1,Gus2}}, although others are available \citep{Davies}. Similarly, the bolometric corrections to the observed M$_{v}$ corresponding to the T$_\mathrm{eff}$ or spectral type can be determined from models. The
total luminosity can also be measured from the star's spectral energy
distribution (SED) based on multi-wavelength photometry when the distance is known \citep{Gordon}.


The red supergiants are the largest stars, but direct measurement of their
angular sizes by interferometry in the past was rare and limited to nearby stars. Thus, most of the radii for the RSGs were estimated from the Stefan--Boltzmann equation with their adopted T$_\mathrm{eff}$  and measured luminosities. The Center for High Angular Resolution Astronomy (CHARA) Array has revolutionized the measurement of stellar diameters including numerous red supergiants using near-infrared interferometry \citep{Ryan}. With the CHARA array, they have also successfully imaged their surfaces, revealing large-scale irregularities, as in AZ Cyg \citep{Ryan2}.


The masses for individual RSGS are most often estimated by comparing their positions on the HRD, their T$_{eff}
$ and L/L$_{\odot}$, with evolutionary tracks. Very few RSGs are in known
binaries with a sufficient baseline to determine mass estimates and radii based on their orbits. See the section below on binarity.  The lower mass limit for the RSGs about 9 M$_{\odot}$ is based on the initial mass for the onset of He-fusion in the non-degenerate core, and the upper limit at about 40 M$_{\odot}$ from the observed upper luminosity boundary on the H-R Diagram, as discussed in Section~\ref{sec2}.

\subsection{Variability}

With their very extended and convective photospheres, it is not surprising that RSGs are variable both photometrically and spectroscopically. The light curves
of most RSGs are classed as either irregular (Ic) or semi-regular (SRc) variables. The observed light curves are as their names imply. The irregular variables
have random, usually small variations with no periodicity, while the light curves of the semi-regular variables can be fit by a period. The SRc variables are fundamental pulsators  and may exhibit more than one recurring period as observed in Betelgeuse.   A period--luminosity relation, with longer periods observed in the most luminous red supergiants, has been reported for some time \citep{Glass,Feast} and is reproduced by pulsation models.


Like their small photometric variations, small changes in the their spectral types, based on the strength of the TiO bands, are observed but are usually on the order of one spectral subtype or less. With their convective surfaces and the increasing evidence from near-infrared interferometry for variable surface features, described in the review by Wittkowski, this is not surprising.  But much larger variations have been observed. HV 11423, in the SMC \citep{Massey}, varied from an early K spectral type to as late as M4 with a corresponding change in its visual magnitude. The authors conclude that the star is highly unstable. The variability may be due to pulsation with an enhanced wind or mass loss.


The red hypergiant and strong OH/IR source VX Sgr is possibly the most interesting and challenging example. The star varied from an M4 Ia to at least as late as M8 corresponding to a large decline in visual brightness by five magnitudes \citep{RMH1972}. Its light curve shows frequent episodes of these large variations but often punctuated with periods of small irregular variability or more limited variations of 1 to 2 magnitudes. VX Sgr is a fundamental mode pulsator. Its variability is more like a MIRA variable or AGB star than an RSG, but its high luminosity is confirmed. VX Sgr is discussed together with other red and yellow hypergiants by Jones and in the review  of masers and molecules in RSGs by Ziurys and Richards in this issue.

\subsection{Binarity}

Few Galactic red supergiants are known binaries. This may initially seem surprising since binarity is common in their massive progenitors, the OB stars, with fractional estimates of 50\% or higher \citep{Sana}. Close binaries that are described as interacting or common envelope binaries  are easily recognized, while others are sufficiently separated that they do not interact and thus evolve independently. In the former situation, the lower mass companion may merge with the RSG as its outer envelope expands, and no longer be detectable. If the two stars are far apart and do not interact, they will have large orbits with long periods measured in years, even decades, and will be  more difficult to recognize.


Antares ($\alpha$ Sco), the best known RSG binary, is an example of the latter type. A visual binary, it has been recognized as a double star since 1819. With a separation of about 2.9~arcsec, at its distance, the two stars have a projected separation of 529 AU.  Although,  with a  period of 1218 yrs(!), its orbit is not well-determined.  The two stars are easily observed with spectral types M1.5 Iab-Ib and B2.5 V and mass estimates of 14 M$_{\odot}$ \citep{Neuhauser} and 7 M$_{\odot}$ \citep{Kud78}, respectively.  The radius of the RSG  primary is estimated at 680 R$_{\odot}$ or 3.2 AU. Many of the uncertainties and range in its published  parameters are attributed to the RSG's variable atmosphere.


Several Galactic RSGs have composite spectra, which is spectra with spectral features  characteristic of stars with different temperatures. 
The most easily recognized are those with a cool star with TiO bands plus a warmer continuum at shorter wavelengths and spectral lines of H or He I typical of an O or B-type star. Further observations are needed though to confirm that the star is a binary and not  a chance superposition in a crowded~field.


The VV Cephei stars are the best examples of a close interacting pair with a red supergiant primary and a  hot  O- or B-type star companion in which the hot secondary  enters the extended atmosphere of the RSG producing ``atmospheric eclipses''. \mbox{See~\citet{Cowley}} for a comprehensive review of the class and a list of known VV Cephei stars with their observed properties. A second group, known as $\zeta$ Aur stars, is similar but  with a warmer K-type primary and a B star secondary. The chromospheric eclipses are important as probes of the red supergiant atmospheres.


VV Cephei itself is a luminous M-type supergiant with a main sequence B star secondary, M2 Iabep + B V,  \citep{Bauer91,Bauer2000} with an orbital period of 20.3 yrs \citep{Wright}. The system has been observed through an entire orbit during and outside its atmospheric eclipse with IUE and HST \citep{Bennett}. Interestingly, the two stars have comparable masses of about  18 M$_{\odot}$ and the RSG has a radius of 5 AU.


Two recent studies have suggested that Betelgeuse's long secondary period is due to an unseen companion \citep{Goldberg,MacLeod}. Betelgeuse's possible binarity and the search for its companion is discussed in the review by Dupree and  Montarg\`{e}s.  Evidence for red supergiant  binarity in nearby galaxies is presented in the Bonanos review of RSGs in nearby galaxies in this Special Issue.

\section{Circumstellar Ejecta and Mass Loss}

Circumstellar dust and mass loss are among the distinguishing characteristics of many evolved massive stars but are foremost for the red supergiants. Today, mass loss, how much mass, and the mass loss mechanisms are critical to understanding their evolution and especially their final stages before the terminal collapse or explosion for massive stars across the H-R Diagram.

\begin{figure}[H]
\includegraphics[width=6.5cm]{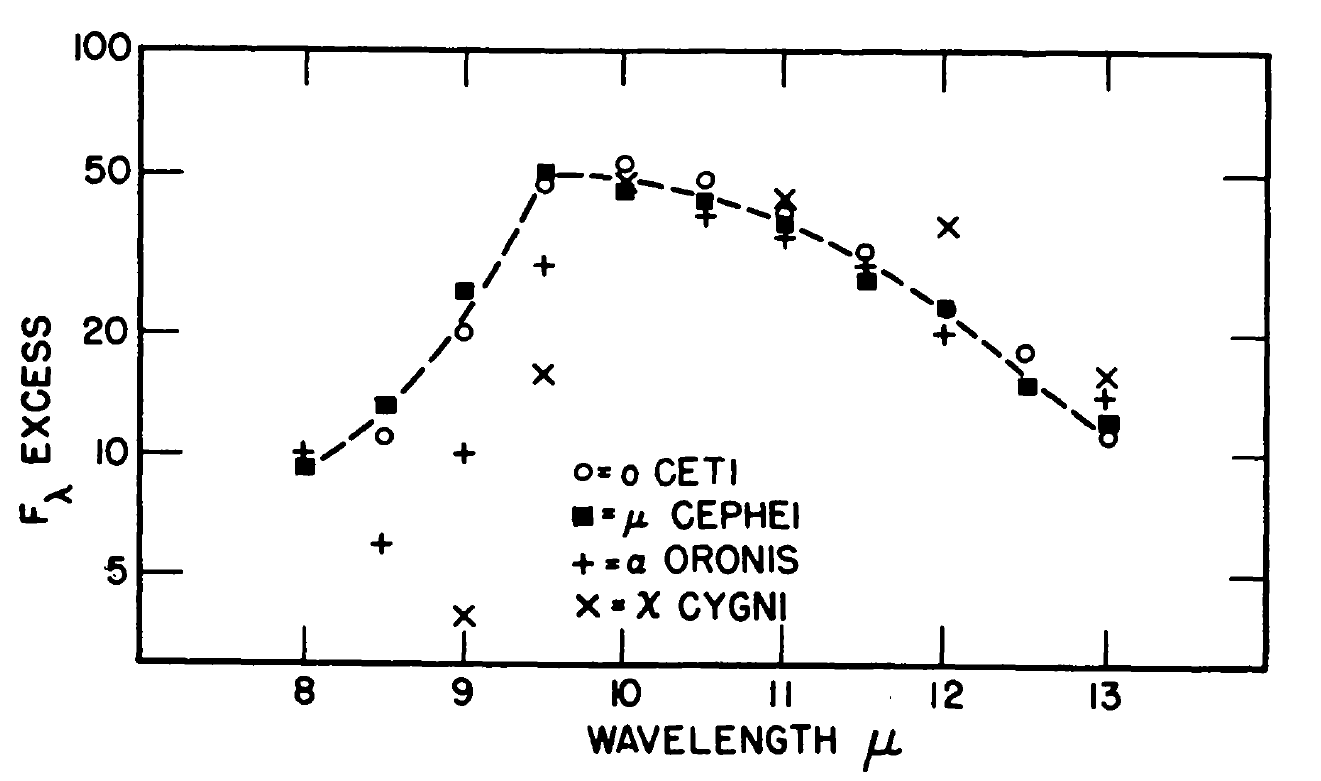}
\includegraphics[width=6.5cm]{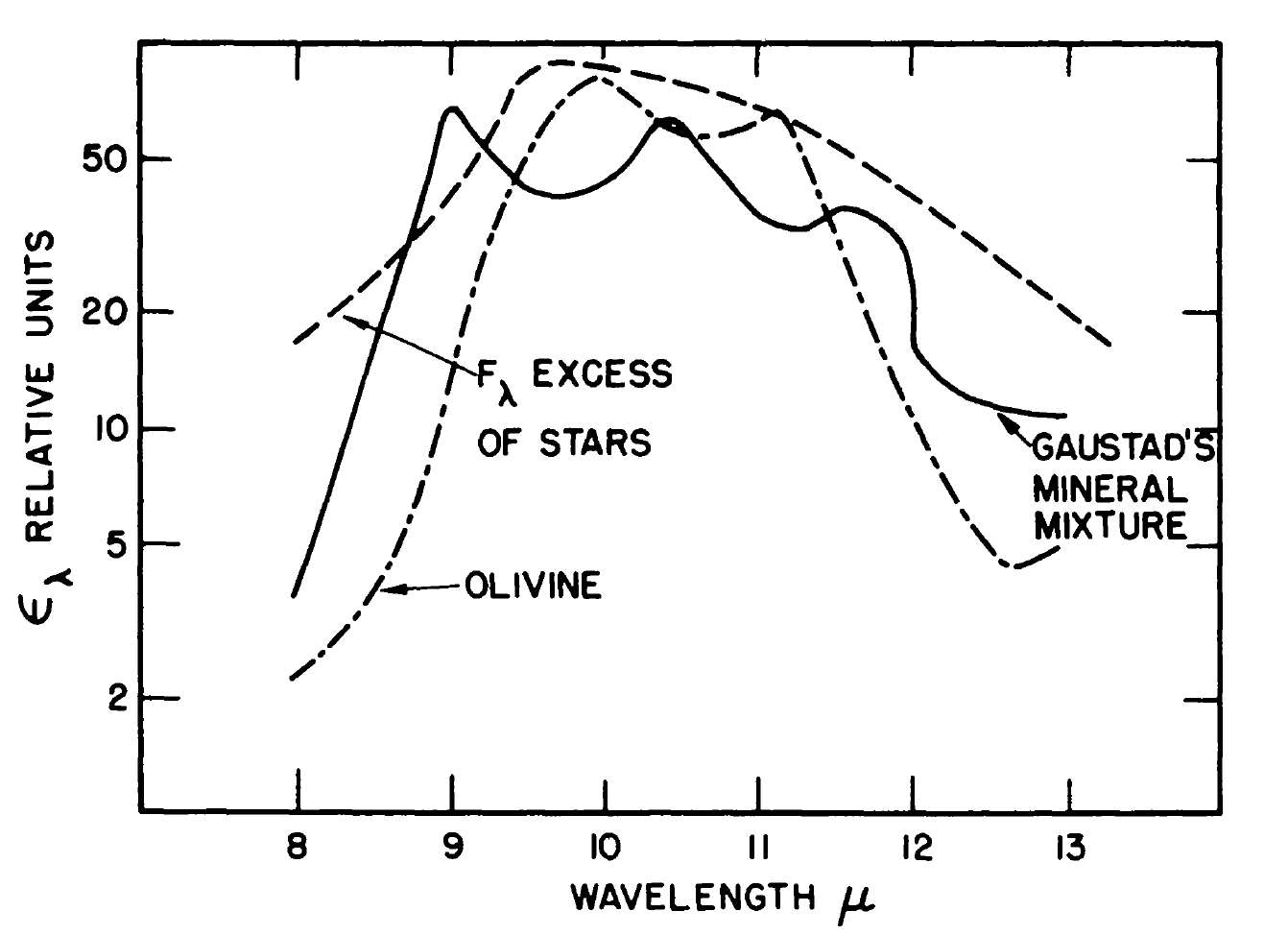}
\caption{(\textbf{Left}): The 10 $\upmu$m emission feature observed in the RSGs $\mu$Cep and $\alpha$ Ori and two late type giants. (\textbf{Right}): Comparison of the emissivity of the 10~$\upmu$m emission (dashed line) with two types of Mg silicates, olivine and a grain mixture,~\citet{Woolf}.\label{fig2}}
\end{figure}

For the red supergiants, the observation of circumstellar dust and mass loss is closely tied to the development of infrared astronomy and the discovery of excess radiation at  wavelengths in the Rayleigh--Jeans tail of their spectral energy distributions (SED).  Infrared astronomy began in the early 1960s with Frank Low’s development of the GaGe bolometers and spread very quickly at several research centers with the detection of infrared emission from several astronomical sources. Interest in infrared astronomy was fueled by the early near-infrared 2~$\upmu$m survey \citep{Neu}. Perhaps the most relevant early discovery was the identification  of the strong 10 to 12 micron emission feature observed in the red supergiants with MgSiO, i.e., silicates, by Woolf and Ney \cite{Woolf}, see Figure~\ref{fig2}. \vspace{-6pt}

\vspace{2mm}

\begin{figure}[H]
\includegraphics[width=6.5cm]{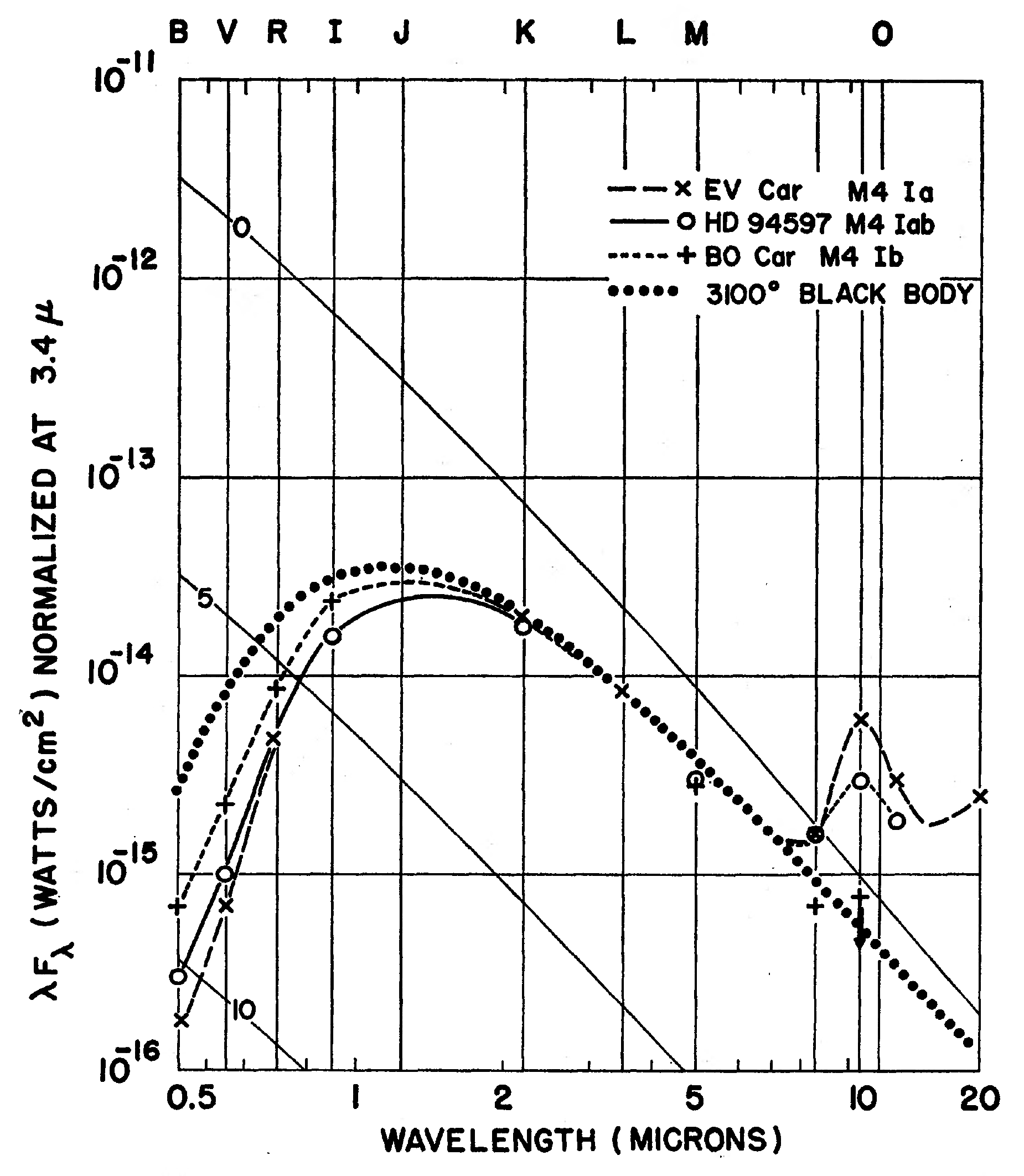}
%
\includegraphics[width=6.5cm]{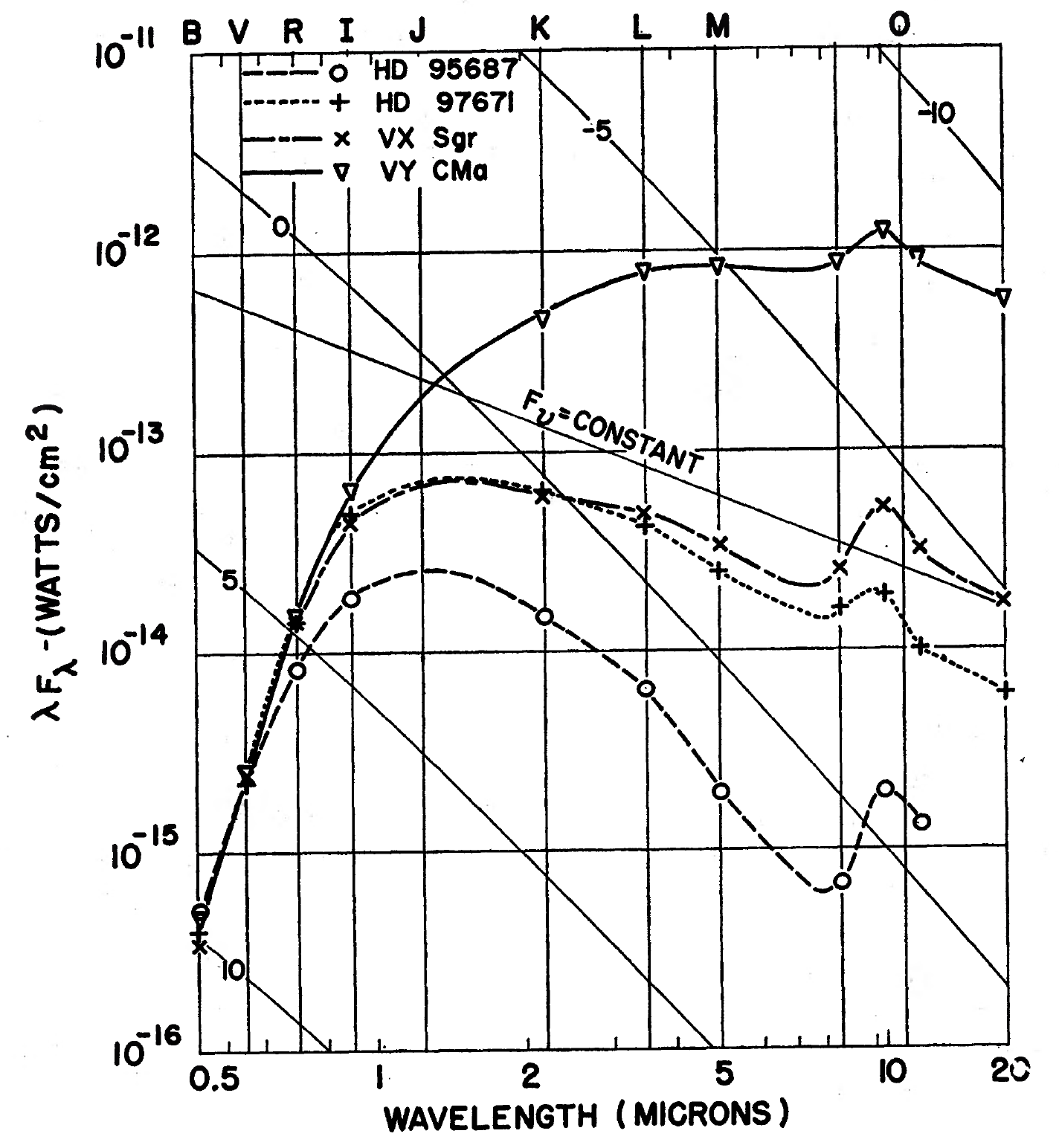}

~~~~~~~~~~~~~~~~~~~~~~~~~~~~~~~~~~~~~~~~~~(\textbf{a}) ~~~~~~~~~~~~~~~~~~~~~~~~~~~~~~~~~~~~~~~~~~~~~~~~~~~~~~~~~~~~~~~~~~~(\textbf{b})

\caption{(\textbf{a}) {The} 
SEDs for three RSGS with the same spectal class but different luminosity classification illustrating the dependence of the 10~$\upmu$m silicate feature on the star's luminosity.  (\textbf{b}) The SEDs for four RSGs. Three of the stars have significant excess radiation in the mid-infrared region from 2 to 8~$\upmu$m in addition to the silicate emission feature indicative of extended circumstellar ejecta \citep{RMH72}.\label{fig3}}
\end{figure}

Red supergiants and other high luminosity cool stars were thus the obvious targets for infrared astronomy in the 1970s. Silicate emission was the distinguishing feature of circumstellar dust associated with red supergiants, but not all
RSGs showed the silicate emission feature. There was a dependence on the temperature and luminosity of the star.  Figure~\ref{fig3}a shows the dependence on luminosity for some early examples. The SEDs of  several of the more luminous RSGS revealed that  excess emission from dust grains  was present at a wide range of wavelengths from 2 to 20~$\upmu$m corresponding to dust temperatures from $\approx$1200 to 300 K and  therefore at a range of distances from the star, Figure~\ref{fig3}b. Two of the  RSGS, VY CMa and VX Sgr in Figure~\ref{fig3}b, are now well-known for their extensive circumstellar dust and high mass loss rates. These stars are often called OH/IR stars for their molecular emission from masers in their ejecta, discussed in the review by Ziurys and Richards in this issue.

\vspace{2mm}

This early work was quickly followed by surveys. The IRAS satellite, launched in 1983, was the first comprehensive infrared survey from space \citep{Neu2}. IRAS added observations at much longer wavelengths, 12, 25, 60, and 100~$\upmu$m, that were not common in the ground-based work. Elias, Frogel, and Humphreys \citep{Elias} combined existing lists of RSGS and candidates in the LMC and SMC with ground-based near-infrared photometry for a comparison with infrared observations of their Galactic counterparts. In this century the ground-based 2MASS near-infrared survey \citep{Skrutski},  mid-infrared observations with IRAC on the Spitzer space telescope, and the WISE \citep{WISE} satellite have extended our knowledge of the RSGS to fainter stars in nearby galaxies.


Mass loss in stars, especially evolved luminous stars of all types, has been known since the 1960s and has been well-observed and studied in both early and late type stars. Mass loss and mass loss rates for the red supergiants are based on their dusty circumstellar environments observed as radiation from grains in the mid-infrared. The mass loss rates are empirically determined, but with assumed grain properties and an uncertain gas to dust ratio of 100 to 200. The most frequently cited works \citep{deJager,loon2005,MJ} show a not surprising dependence on
the luminosity of the star as suggested in early work (Figure~\ref{fig3}a). Vink and Subhahit \citep{Vink} suggested that $\dot{M}$ depends on L/M, the Eddington factor. The stars with the highest $\dot{M}$ are also closest to the Eddington limit for their mass.

\begin{figure}[H]
\includegraphics[width=0.98\textwidth]{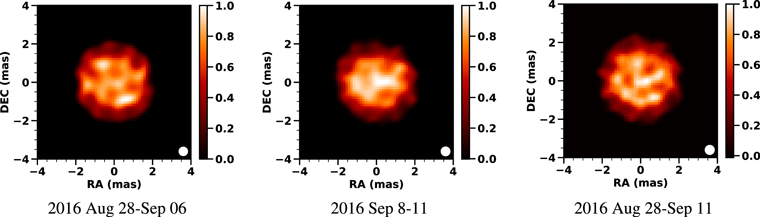}
\caption{{CHARA} 
near-infrared interferometric images of AZ Cyg illustrating the variability of the surface asymmetries \citep{Ryan2}.\label{fig4}}
\end{figure}

The three leading methods for red supergiant mass loss are radiation pressure on grains, pulsation, and convection or some combination. But there are problems, especially explaining the high mass loss measured for some of most luminous RSG with their visible ejecta and outflows. Radiation pressure on grains works well for the AGB stars \citep{Kwok,loon2006}, with fundamental mode pulsation, which enhances the mass loss driven by the dust grains. However, in the less variable RSGs with very extended atmospheres,  radiation pressure and pulsation are not adequate to elevate the material to the dust and molecular formation zones. The presence and variability of large-scale asymmetries on the surfaces  of many RSGs such as Betelgeuse \citep{Gilliland}, AZ Cyg (Figure~\ref{fig4}) and RW Cep \citep{Anugu,Anugu2} lends support for convection as important.

The  recent  $\dot{M}$ vs. luminosity relations for the Magellanic Clouds \citep{Yang,Anton} plus the RSG-rich clusters in the Milky Way \citep{RMH2020} are curved and show an  upward turn for the most luminous red supergiants. These results plus the high mass loss episodes for the same stars suggest that an additional mechanism is involved; not just convection, but active surfaces.  Massive directed outflows that result in a dimming of the star are now reported for the more typical red supergiant Betelgeuse \citep{Dupree}, the yellow hypergiant RW Cep, and in the historical light curve for the red hypergiant VY CMa \citep{RMH2021,RMH2024}. The masses of some of the outflows, the knots and condensations, in the ejecta of VY CMa are estimated at \linebreak$\ge$0.02 M$_{\odot}$. These massive outflows contribute significantly to their high $\dot{M}$ \citep{HJ} and  may be comparable to coronal mass ejections with magnetic fields. The presence of magnetic fields is  supported by the circular polarization and Zeeman effect in the masers in the ejecta of RSGS like VY CMa, VX Sgr, and S Per \citep{Vlemmings}.


Mass loss and the circumstellar environments of the RSGs are discussed in more detail in the reviews by van Loon and by Wittkowski in this Special Issue.

\section{The Final State---Supernovae}

Fritz Zwicky's famous classification \citep{Zwicky} of observed supernovae separated them into five types based primarily on their spectral characteristics. The familiar types I and II were already recognized.  Zwicky credits Rudolph Minkowski \citep{Minkowski} with first recognizing the second type, Type II, with a strong blue continuum and Hydrogen Balmer emission lines not seen in the Type I spectra.  Zwicky's Types III and IV were represented by only a few supernovae which today would probably be called Type IIpec. His subluminous Type V examples are probably due to high mass loss eruptions from very massive stars, the ``supernova impostors''.




\textls[-15]{The first references suggesting that RSGs are Type II supernova progenitors are \mbox{\citet{Barbon}} who suggested a red supergiant with 10 M$_{\odot}$, followed by \mbox{ \citet{Barbaro}} who discuss the pre-supernova stage and conclude that the star must have a large extended envelope and be at least 10--14 M$_{\odot}$, which they emphasize would be as a red supergiant. Subsequent papers in the next decade discuss the red supergiants as progenitors of the Type II SNe. Barbon et al. \citep{Barbon79} recognized the division of the Type II class based on the shape of their light curves: 1. those with a flat plateau at constant luminosity lasting several weeks, Type II-P,  and 2. those with a linear decline in luminosity, Type II-L. With their very extended and massive  atmospheres, the RSGs are the progenitors of the Type II-P supernovae.}


Perhaps the early work most relevant to the discussions in the reviews here is the paper by Maeder \citep{Maeder} on the pre-supernova state of RSGs, the role of mass loss and post-RSG evolution. He showed that with high mass loss, the H, C, O core reaches a critical mass and the star transitions back to a warmer temperature.


The most interesting and challenging recent development with respect to the final state of the RSGS is the ``red supergiant problem''.  A survey of the Type II-P \mbox{progenitors \citep{Smartt2009,Smartt2015}} revealed an upper  limit to the initial masses of only 17 M$_{\odot}$ for their likely RSG progenitors. Of course, many red supergiants are known with greater initial masses, up to 30 or more  M$_{\odot}$. ~\citet{Smartt2009} suggested two possible explanations; underestimated luminosities (and masses) due to improper extinction corrections, or an alternative to core collapse as a Type II-P supernova.  The latter could be direct collapse to the black hole or post-RSG evolution to a warmer state followed by a  terminal explosion.


The red supergiant progenitors of the Type II-P supernovae is reviewed in this Special Issue by van Dyk with well-studied recent examples.

\section{Current Questions---The Red Supergiant Problem, Episodic Mass Loss, and Post Red Supergiants}

The lack or dearth of RSG supernova progenitors above about 20 M$_{\odot}$ and the high mass loss episodes in the most luminous RSGs  are two outstanding current questions for red supergiant evolution and their final terminal state. Indeed, the two are probably related.


\textls[-5]{{\it {The Red Supergiant Problem:} 
} Some authors have questioned Smartt's \mbox{conclusions \citep{Beasor1,Beasor2}}, arguing that higher extinction due to dusty circumstellar ejecta results in an underestimate of the luminosity and mass.  Although more dusty environments can alter the SED, the amount of extinction is not sufficient to solve the red supergiant problem  for stars above \linebreak$\approx$20 M$_{\odot}$ \citep{Walmswell,Kilpatrick}.}

Many of the most luminous RSGs, those most likely with initial masses above 20~M$_{\odot},$ have high mass loss rates on the order of 10$^{-5}$ to more than 10$^{-4}$ M$_{\odot}$ per year. With these high  $\dot{M}$, the central core may reach a critical mass, about two-thirds of the total mass~\citep{Maeder}, to send the stars on a blue-loop to warmer temperatures. Observations of the  yellow hypergiants with extended dusty environments and high  $\dot{M}$ may be evidence for post-red supergiant evolution. The yellow and red hypergiants are discussed in the review by Jones in this Special Issue, and  the interior evolution of evolved massive stars and their evolutionary tracks are described in the review by Ekstrom and Georgy.



The other possibility suggested by Smartt, is a direct collapse to a black hole without the explosive signature of a supernova, i.e., a failed supernova. A search for failed supernovae has identified a  candidate, N6946-BH1 \citep{Gerke,Adams}, which brightened briefly to $\sim$10$^{6}$ L$_{\odot}$ before fading below its pre-outburst luminosity and with an estimated initial mass of about 25 M$_{\odot}$. Based on its SED and warm wind, \citep{RH-BH1} suggested the progenitor may have been a yellow hypergiant in a post-RSG state.  The rarity of these events suggests that they may not be the explanation for the red supergiant problem, although a  second candidate for a failed SNe has been recently reported in M31 \citep{De}.


{\it {High Mass Loss Episodes and Post-Red Supergiant Evolution:}} The very high mass loss rates measured for the red hypergiants can result in post-red supergiant evolution to warmer temperatures where they may still end their lives as a supernova. The origin of their high mass loss is still uncertain. With the large and variable asymmetries or hot-spots now observed on many RSGs, a likely explanation for the high mass loss episodes is large-scale surface activity accompanied by massive outflows.

We would expect to observe these post RSG stars probably as yellow hypergiants with a record of their high mass loss history and on-going mass loss and instabilities.
Several yellow hypergiants exhibit short-term variations in their apparent spectral types or surface temperatures. These apparent transits on the HRD are attributed to pulsational instabilities or changes in the wind, not evolution in the interior resulting in structural changes from a red supergiant to a warmer star which are expected to take a few thousand years. Var A in M33, however, has been observed to transit from yellow hypergiant to an apparent RSG state lasting decades, due to a high mass loss event, and back to its warmer state \citep{RMH06}.

\begin{figure}[H]
\includegraphics[width=4.2cm]{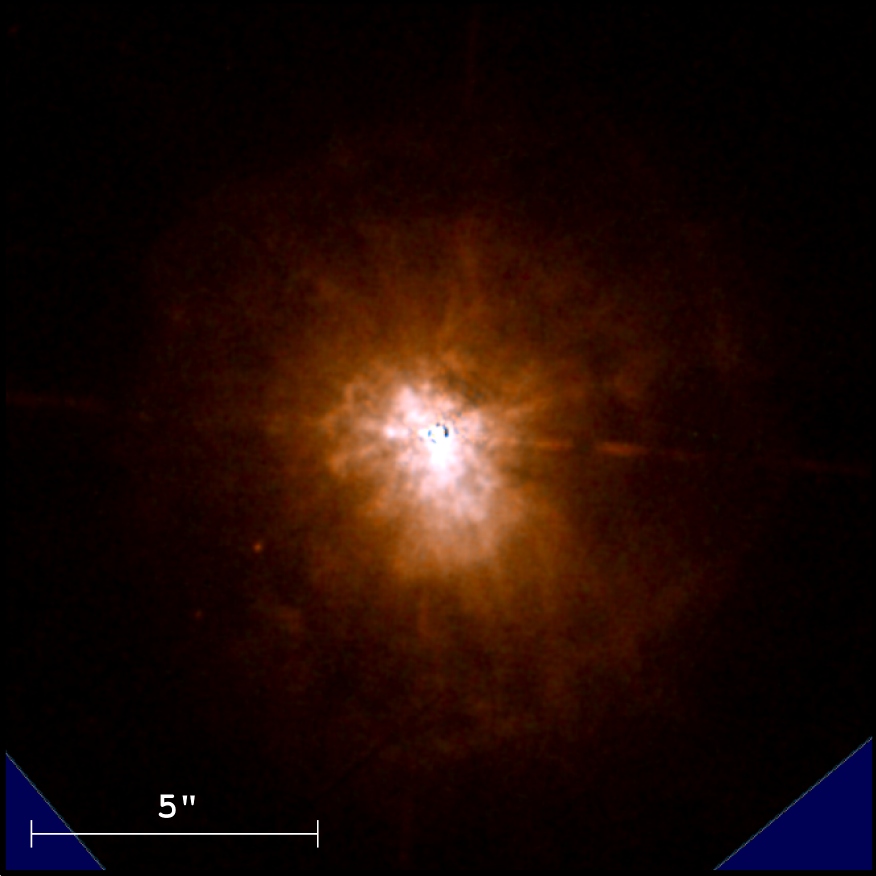}
\includegraphics[width=4.2cm]{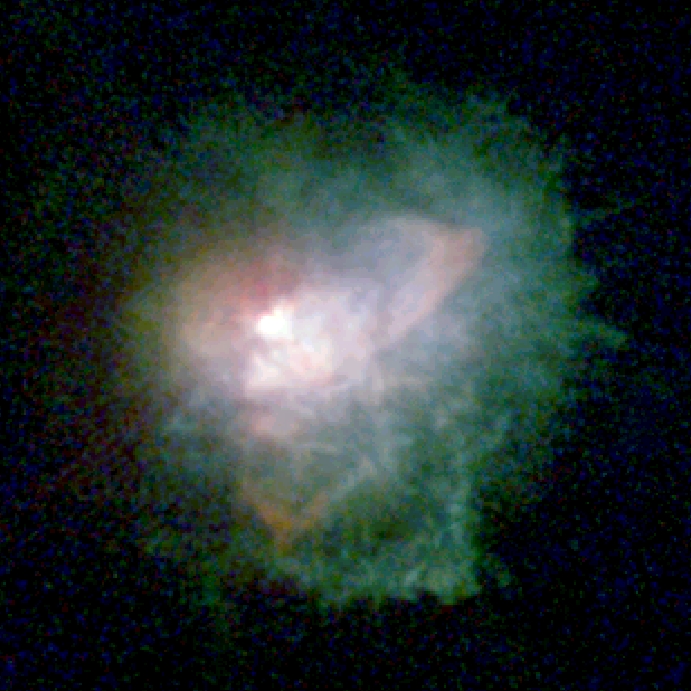}
\includegraphics[width=4.2cm]{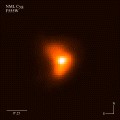}
\caption{\textls[-15]{{Multiwavelength} 
HST WFPC2 images of the hypergiants (\textbf{Left}): IRC+10420 \citep{RMH1997}, (\textbf{Center}):~VY CMa \citep{Smith}, (\textbf{Right}): NML Cyg \citep{Schuster}.}\label{fig5}}
\end{figure}

\textls[-15]{The yellow hypergiant IRC+10420 may be the best example of post-red supergiant evolution. Its HST image (Figure~\ref{fig5}) shows its record of high mass loss episodes with numerous knots, small semi-circular arcs, and more distant arcs or shells. Tiffany et al. \citep{Tiffany} found that ejecta at distances up to 8--9 arc sec from the star were ejected about 5000~yrs ago. In an independent  study Shenoy et al. \citep{Shenoy} showed that its long-wavelength SED could best be fit by two epochs of high mass loss; one lasting from 6000 up to 2000 yrs ago. These observations support a prior high mass loss state for IRC+10420 probably as a red~supergiant. }

\textls[-20]{The final fate of the post RSG yellow hypergiants is not known. Very few yellow hypergiants have been confirmed as supernova progenitors. The best example is \mbox{SN2011dh \citep{Maund,vandyk11,vandyk13}}, a Type IIb supernova identified with a yellow supergiant. If the star continues to lose mass and evolve to even warmer temperatures, shedding its outer envelope, it may be seen eventually as a Luminous Blue Variable (LBV) or Wolf Rayet star. }




Another possibility, not often discussed, is whether some RSGs, like the extreme red hypergiants VY CMa and NML Cyg (Figure~\ref{fig5}), may be second-generation RSGs, After their post-red supergiant blue transit on the HRD, they return to the red supergiant region, similar to lower mass stars and are now in a very short high mass loss state prior to the terminal explosion or collapse to a black hole.
\vspace{6pt}

\funding{{ This research received no external funding}}

\conflictsofinterest{{The authors declare no conflict of interest. }}


\abbreviations{Abbreviations}{
The following abbreviations are used in this manuscript:\\

\noindent
\begin{tabular}{@{}ll}
HRD & Hertzsprung--Russell Diagram, H-R Diagram\\
RSG & red supergiant \\
SED & spectral energy distribution \\
\end{tabular}
}

\reftitle{References}
\begin{adjustwidth}{-\extralength}{0cm}

%
\PublishersNote{}
\end{adjustwidth}

\begin{thebibliography}{999}
\bibitem[Levesque(2017)]{EL}
Levesque, E.  {\em The Astrophysics of Red Supergiants}; Institute of Physics Publishing: Bristol, UK, {2017}; ISBN 978-0-7503-1330-8.

\bibitem[Martin \& Humphreys(2023)]{JM}
Martin, J.C.; Humphreys, R.M. A Census of the Most Luminous Stars. I. The Upper HR Diagram for the Large Magellanic Cloud. {\em Astron. J.} {\bf 2023}, {\em 166},  214--225. [\href{http://doi.org/10.3847/1538-3881/ad011e}{CrossRef}]

\bibitem[Bidelman(1947)]{WPB}
Bidelman, W.P. The M-Type Supergiant Members of the Double Cluster in Perseus. {\em Astrophys. J.} {\textbf{1947}}, {\emph{105}}, 492--496. [\href{http://dx.doi.org/10.1086/144923}{CrossRef}]

\bibitem[Humphreys \& Davidson(1979)]{HD79}
Humphreys, R.M.; Davidson, K. Studies of Luminous stars in Nearby Galaxies.
III---Comments on the Evolution of the Most Massive Stars in the Milky Way
and the Large Magellanic Cloud. {\em Astrophys. J.} {\bf 1979}, {\em 232}, 409--420. [\href{http://dx.doi.org/10.1086/157301}{CrossRef}]


\bibitem[McCarthy(1950)]{McCarthy}
McCarthy, M.F. Fr. Secchi and Stellar Spectra. {\em Pop. Astron.} {\bf 1950}, {\em 58}, 153--168.

\bibitem[Cannon \& Pickering(1918-1924)]{HA}
Cannon, A.J.;  Pickering, E.C.  The Henry Draper Catalog. {\em {HA}
} {\bf 1918--1924}, {\em 91--99}.


\bibitem[Morgan \& Keenan(1943)]{MK}
Morgan, W.W.; Keenan, P.C.; Kellman, E. {\em An Atlas of Stellar Spectra, with an Outline of Spectral Classification}; University of Chicago Press: Chicago, IL, USA, 1943.

\bibitem[Flower(1977)]{Flower}
Flower, P.J. Transformations from Theoretical H-R Diagrams to C-M Diagrams. {\em Astron. Astrophys.} {\bf 1977}, {\em 54}, 31--39.

\bibitem[Flower.(1996)]{Flower2}
Flower, P.J. Transformations from Theoretical Hertzsprung-Russell Diagrams to Color-Magnitude Diagrams: Effective Temperatures, B-V Colors, and Bolometric Corrections. {\em Astrophys. J.} {\bf 1996}, {\em 469}, 355--365. [\href{http://dx.doi.org/10.1086/177785}{CrossRef}]

\bibitem[Levesque, et al.(2005)]{Levesque1}
Levesque, E.M.; Massey, P.; Olsen, K.A.G.; Plez, B.; Josselin, E.; Maeder, A.; Meynet, G. The Effective Temperature Scale of Galactic Red Supergiants: Cool, but Not as Cool as We Thought. {\em Astrophys. J.} {\bf 2005}, {\em 628}, 973--985. [\href{http://dx.doi.org/10.1086/430901}{CrossRef}]

\bibitem[Levesque, et al.(2006)]{Levesque2}
Levesque, E.M.; Massey, P.; Olsen, K.A.G.; Plez, B.; Meynet, G.; Maeder, A.  The Effective Temperatures and Physical Properties of Magellanic Cloud Red Supergiants: The Effects of Metallicity.   {\em Astrophys. J.} {\bf 2006}, {\em 645}, 1102--1117. [\href{http://dx.doi.org/10.1086/504417}{CrossRef}]

\bibitem[Plez, et al.(1992)]{Plez}
Plez, B.; Brett, J.M.; Nordland, A. Spherical opacity sampling model atmospheres for M-giants. I. Techniques, data and discussion. {\em Astron. Astrophys.} {\bf 1992}, {\em 258}, 551--571.

\bibitem[Gustafsson, et al.(2003)]{Gus1}
Gustafsson, B.; Edvardsson, B.; Eriksson, K.; Mizuno-Wiedner, M.; Jørgensen, U.G.;
Plez, B. A Grid of Model Atmospheres for Cool Stars.  In Proceedings of the ASP Conference Proceedings, {Tuebingen, Germany, 8--12 April 2002}
; Hubeny, I., Mihalas, D., Werner, K., Eds.; Stellar Atmosphere Modeling; Astronomical Society of the Pacific: San Francisco, CA, USA, 2003; pp. 331--334.

\bibitem[Gustafsson, et al.(2008)]{Gus2}
Gustafsson, B.; Edvardsson, B.; Eriksson, K.; Jørgensen, U.G.; Nordlund, Å.; Plez, B. A grid of MARCS model atmospheres for late-type stars. I. Methods and general properties. {\em Astron. Astrophys.} {\bf 2008}, {\em 486},  951--970. [\href{http://dx.doi.org/10.1051/0004-6361:200809724}{CrossRef}]

\bibitem[Davies, et al.(2013)]{Davies}
Davies, B.; Kudritzki, R.-P.; Plez, B.; Trager, S.; Lançon, A.; Gazak, Z.; Bergemann, M.; Evans, C.; Chiavassa, A. The Temperatures of Red Supergiants. {\em Astrophys. J.} {\bf 2013}, {\em 767}, 3--23. [\href{http://dx.doi.org/10.1088/0004-637X/767/1/3}{CrossRef}]

\bibitem[Gordon et al.(2016)]{Gordon}
Gordon, M.S.; Humphreys, R.M.; Jones, T.J. Luminous and Variable Stars in M31 and M33: The Yellow and Red Supergiants and Post=Red Supergiant Evolution. {\em Astrophys. J.} {\bf 2016}, {\em 825}, 50--67. [\href{http://dx.doi.org/10.3847/0004-637X/825/1/50}{CrossRef}]

\bibitem[Norris et al.(2023)]{Ryan}
Norris, R.P.; Beltran, Y.; Lucero, K.; Frothingham, D. An Optical Interferometric Look at Red Supergiants. {\em Bull. AAS}  {\bf 2023}, {\em 55},~{163.11}.

\bibitem[Norris et al.(2021)]{Ryan2}
Norris, R.P.; Baron, F.R.; Monnier, J.D.; Paladini, C.; Anderson, M.D.; Martinez, A.O.; Schaefer, G.H.; Che, X.; Chiavassa, A.; Connelley, M.S.; et al. Long Term Evolution of Surface Features on the Red Supergiant AZ Cyg. {\em Astrophys. J.} {\bf 2021}, {\em 919}, 124--145. [\href{http://dx.doi.org/10.3847/1538-4357/ac0c7e}{CrossRef}]

\bibitem[Glass(1979)]{Glass}
Glass, I.S. Infrared observations of late-type supergiants in the Magellanic Clouds. {\em Mon. Not. R. Astron. Soc.} {\bf 1979}, {186}, 317--326. [\href{http://dx.doi.org/10.1093/mnras/186.2.317}{CrossRef}]

\bibitem[Feast et al.(1980)]{Feast}
Feast, M.W.;  Catchpole, R.M.; Carter, B.S.; Roberts, G. A period-luminosity relation for supergiant red variables in the LMC. {\em Mon. Not. R. Astron. Soc.} {\bf 1980}, {\em193}, 377--380. [\href{http://dx.doi.org/10.1093/mnras/193.2.377}{CrossRef}]

\bibitem[Massey et al.(2007)]{Massey}
Massey, P.; Levesque, E.M.; Olsen, K.A.G.; Plez, B.; Skiff, B.A. HV 11423: The Coolest Supergiant in the SMC. {\em Astrophys. J.} {\bf 2007}, {\em 660}, 301--310. [\href{http://dx.doi.org/10.1086/513182}{CrossRef}]

\bibitem[Humphreys \& Lockwood(1972)]{RMH1972}
Humphreys, R.M.; Lockwood, G.W. Spectroscopic and Photometric Changes in the Peculiar Infrared Star VX Sagittarius. {\em Astrophys. J.} {\bf 1972}, {\em 172}, L59--L62. [\href{http://dx.doi.org/10.1086/180891}{CrossRef}]

\bibitem[Sana et al.(2012)]{Sana}
Sana, H.; de Mink, S.E.; de Koter, A.; Langer, N.; Evans, C.J.; Gieles, M.; Gosset, E.; Izzard, R.G.; Le Bouquin, J.-B.; Schneider, F.R.N. Binary Interaction Dominates the Evolution of Massive Stars. {\em Science} {\bf 2012}, {\em 337}, 444--453. [\href{http://dx.doi.org/10.1126/science.1223344}{CrossRef}]

\bibitem[Neuhauser et al.(2022)]{Neuhauser}
Neuhäuser, R.; Torres, G.; Mugrauer, M.; Neuhäuser, D.L.; Chapman, J.; Luge, D.; Cosci, M.  Colour evolution of Betelgeuse and Antares over two millennia, derived from historical records, as a new constraint on mass and age. {\em Mon. Not. R. Astron. Soc.} {\bf 2022}, {\em 516}, 693--719. [\href{http://dx.doi.org/10.1093/mnras/stac1969}{CrossRef}]

\bibitem[Kudritzki \& Reimers(1978)]{Kud78}
Kudritzki, R.P.; Reimers, D. On the absolute scale of mass-loss in red giants. II. Circumstellar absorption lines in the spectrum of alpha Sco B and mass-loss of alpha Sco A.  {\em Astron. Astrophys.} {\bf 1978}, {\em 70}, 227--239.

\bibitem[Cowley(1969)]{Cowley}
Cowley, A.P. The VV Cephei Stars. {\em Publ. Astron. Soc. Pac.} {\bf 1969}, {\em 81}, 297--331. [\href{http://dx.doi.org/10.1086/128784}{CrossRef}]

\bibitem[Bauer et al.(1991)]{Bauer91}
Bauer, W.H.; Stencel, R.E.; Neff, D.H. Twelve years of IUE spectra of the interacting binary VV Cephei. {\em Astron. Astrophys. Suppl. Ser.} {\bf 1991}, {\em 90}, 175--183.

\bibitem[Bauer \& Bennett(2000)]{Bauer2000}
Bauer, W.H.; Bennett, P.D. The Ultraviolet Spectrum of VV Cephei Out of Eclipse. {\em Publ. Astron. Soc. Pac.} {\bf 2000}, {\em 112}, 31--49. [\href{http://dx.doi.org/10.1086/316479}{CrossRef}]

\bibitem[Wright(1977)]{Wright}
Wright, K.O. The System of VV Cephei Derived from an Analysis of the H-alpha Line. {\em J. R. Astron. Soc. Can.} {\bf 1977},  {\em 71}, 152--193.

\bibitem[Bennett \& Bauer(2015)]{Bennett}
Bennett, P.D.; Bauer, W.H. The Special Case of VV Cephei. In {\em
Giants of Eclipse: The $\zeta$ Aurigae Stars and Other Binary Systems}; Ake,~T.B., Griffin, E., Eds.; Springer International Publishing: Cham, Switzerland,  2015; p. 85, ISBN 978-3-319-09197-6.

\bibitem[Goldberg et al. (2024]{Goldberg}
Goldberg, J.A.; Joyce, M.; Molnar, L.A. A Buddy for Betelgeuse: Binarity as the Origin of the Long Secondary Period in {$\alpha$} Orionis. {\em Astrophys. J.} {\bf 2024}, {\em 977}, 35--55. [\href{http://dx.doi.org/10.3847/1538-4357/ad87f4}{CrossRef}]

\bibitem[MacLeod et al.(2025)]{MacLeod}
MacLeod, M.; Blunt, S.; De Rosa, R.J.; Dupree, A.K.; Granzer, T.; Harper, G.M.; Huang, C.D.; Leiner, E.M.; Loeb, A.; Nielsen, E.L.; et al. Radial Velocity and Astrometric Evidence for a Close Companion to Betelgeuse. {\em Astrophys. J.} {\bf 2025}, {\em 978}, 50--81. [\href{http://dx.doi.org/10.3847/1538-4357/ad93c8}{CrossRef}]


\bibitem[Neugebauer \& Leighton(1969)]{Neu}
Neugebauer, G.; Leighton, R.B. {\em Two-Micron Sky Survey, a Preliminary Catalog}; NASA Special Publication: {Washington, DC, USA}, 1969.

\bibitem[Woolf \& Ney(1969)]{Woolf}
Woolf, N.J.; Ney, E.P. Circumstellar Infrared Emission from Cool Stars. {\em Astrophys. J.} {\bf 1969}, {\em 155}, 181--184. [\href{http://dx.doi.org/10.1086/180331}{CrossRef}]

\bibitem[Humphreyset al.(1972)]{RMH72}
Humphreys, R.M.; Strecker, D.W.; Ney, E.P. Spectroscopic and Photometric Observations of M Supergiants in Carina. {\em Astrophys. J.} {\bf 1972}, {\em 172}, 75--88. [\href{http://dx.doi.org/10.1086/151329}{CrossRef}]

\bibitem[Neugebauer et al(1983)]{Neu2}
Neugebauer, G.; Habing, H.J.; van Duinen, R.; Aumann, H.H.; Baud, B.; Beichman, C.A.; Beintema, D.A.; Boggess, N.; Clegg, P.E.; de Jong, T.; et al. The Infrared Astronomical Satellite (IRAS) mission. {\em Astrophys. J.} {\bf 1984}, {\em 278}, L1--L6. [\href{http://dx.doi.org/10.1086/184209}{CrossRef}]

\bibitem[Elias, Frogel, \& Humphreys(1985)]{Elias}
Elias, J.H.; Frogel, J.A.; Humphreys, R.M. M supergiants in the Milky Way
and the Magellanic Clouds: Colors, spectral types, and luminosities. {\em Astrophys. J. Suppl. Ser.}
{\bf 1985}, {\em 57}, 57--131. [\href{http://dx.doi.org/10.1086/190997}{CrossRef}]

\bibitem[Skrutskie et al.(2006)]{Skrutski}
Skrutskie, M.F.; Cutri, R.M.; Stiening, R.; Weinberg, M.D.; Schneider, S.; Carpenter, J.M.; Beichman, C.; Capps, R.; Chester, T.; Elias, J.; et al. The Two Micron All Sky Survey (2MASS). {\em Astron. J.} {\bf 2006}, {\em 131}, 1163--1183.

\bibitem[Blain et al.(2005)]{WISE}
Blain, A.; Lonsdale, C.; Cutri, R.; WISE Team. The WISE mission: An Atlas for Luminous Galaxy Evolution. {\em  Bull.  Am. Astron. Soc.} {\bf 2005}, {\em 37},  1393.

\bibitem[deJager et al.(1988)]{deJager}
de Jager, C.; Nieuwenhuijzen, H.; van der Hucht, K.A. Mass Loss Rates in the Hertzsprung-Russell Diagram. {\em  Astron. Astrophys. Suppl. Ser.} {\bf 1988}, {\em 72}, 259--289.


\bibitem[van Loon et al.(2005)]{loon2005}
van Loon, J.T.; Cioni, M.-R.L.; Zijlstra, A.A.; Loup, C. An Empirical Formula for the Mass-Loss Rates of Dust-Enshrouded Red Supergiants and Oxygen-Rich Asymptotic Giant Branch Stars. {\em Astron. Astrophys.} {\bf 2005}, {\em 438}, 273--289. [\href{http://dx.doi.org/10.1051/0004-6361:20042555}{CrossRef}]

\bibitem[Mauron \& Josselin(2011)]{MJ}
Mauron, N.; Josselin, E.  The Mass-Loss Rates of Red Supergiants and the de Jager Prescription. {\em Astron. Astrophys.} {\bf 2011}, {\em 526}, 156--170. [\href{http://dx.doi.org/10.1051/0004-6361/201013993}{CrossRef}]

\bibitem[Vink \& Subhahit(2023)]{Vink}
Vink, J.S.; Subhahit, G.N.  Exploring the Red Supergiant wind kink. A Universal mass-loss concept for massive stars. {\em Astron. Astrophys.} {\bf 2023}, {\em 678}, L3--L7. [\href{http://dx.doi.org/10.1051/0004-6361/202347801}{CrossRef}]

\bibitem[Kwok(1975)]{Kwok}
Kwok, S. Radiation Pressure on Grains as a Mechanism for Mass Loss in Red Giants. {\em Astrophys. J.} {\bf 1975}, {\em 198}, 583--591. [\href{http://dx.doi.org/10.1086/153637}{CrossRef}]

\bibitem[van Loon(2006)]{loon2006}
van Loon, J.T.; Marshall, J.R.; Cohen, M.; Matsuura, M.; Wood, P.R.; Yamamura, I.; Zijlstra, A.A. Very Large Telescope three micron spectra of dust-enshrouded red giants in the Large Magellanic Cloud. {\em Astron. Astrophys.} {\bf 2006}, {\em 447}, 971--989. [\href{http://dx.doi.org/10.1051/0004-6361:20054222}{CrossRef}]

\bibitem[Gilliland \& Dupree(1996)]{Gilliland}
Gilliland, R.L.; Dupree, A.K. First Image of the Surface of a Star with the Hubble Space Telescope.  {\em Astrophys. J.} {\bf 1996}, {\em 463}, L29--L32. [\href{http://dx.doi.org/10.1086/310043}{CrossRef}]


\bibitem[Anugu et al.(2023)]{Anugu}
Anugu, N.; Baron, F.; Gies, D.R.; Lanthermann, C.; Schaefer, G.H.; Shepard, K.A.; Brummelaar, T.T.; Monnier, J.D.; Kraus, S.;\linebreak Le Bouquin, J.-B.; et al. The Great Dimming of the Hypergiant Star RW Cephei: CHARA Array Images and Spectral Analysis. {\em Astron. J.} {\bf 2023}, {\em 166}, 78--89. [\href{http://dx.doi.org/10.3847/1538-3881/ace59d}{CrossRef}]

\bibitem[Anugu et al.(2024)]{Anugu2}
Anugu, N.; Gies, D.R.; Roettenbacher, R.M.; Monnier, J.D.; Montargés, M.; Mérand, A.; Baron, F.; Schaefer, G.H.; Shepard, K.A.; Kraus, S.; et al. Time Evolution Images of the Hypergiant RW Cephei during the Rebrightening Phase Following the Great Dimming. {\em Astrophys. J.} {\bf 2024}, {\em 973}, L5--L24. [\href{http://dx.doi.org/10.3847/2041-8213/ad736c}{CrossRef}]

\bibitem[Yang et al.(2023)]{Yang}
Yang, M.; Bonanos, A.Z.; Jiang, B.; Zapartas, E.; Gao, J.; Ren, Y.; Lam, M.I.; Wang, T.; Maravelias, G.; Gavras, P.; et al. Evolved Massive Stars at Low-metallicity. V. Mass-loss Rate of Red Supergiant Stars in the Small Magellanic Cloud. {\em  Astron. Astrophys.} {\bf 2023}, {\em 676}, 84--100. [\href{http://dx.doi.org/10.1051/0004-6361/202244770}{CrossRef}]


\bibitem[Antoniadis et al.(2024)]{Anton}
Antoniadis, K.; Bonanos, A.; de Wit, S.; Zapartas, E.; Munoz-Sanchez, G.; Maravelias, G. Establishing a Mass-loss rate Relation for Red Supergiants in the Large Magellanic Cloud. {\em Astron. Astrophys.} {\bf 2024}, {\em 686}, 88--105. [\href{http://dx.doi.org/10.1051/0004-6361/202449383}{CrossRef}]


\bibitem[Humphreys et al.(2020)]{RMH2020}
Humphreys, R.M.; Helmel, G.; Jones, T.J.; Gordon, M.S. Exploring the Mass-loss Histories of the Red Supergiants. {\em Astron. J.} {\bf 2020}, {\em 160}, 145--165;. [\href{http://dx.doi.org/10.3847/1538-3881/abab15}{CrossRef}]

\bibitem[Dupree et al.(2020)]{Dupree}
Dupree, A.K.; Strassmeier, K.G.; Matthews, L.D.; Uitenbroek, H.; Calderwood, T.; Granzer, T.; Guinan, E.F.; Leike, R.; Montargès,~M.; Richards, A.M.S.; et al. Spatially Resolved Ultraviolet Spectroscopy of the Great Dimming of Betelgeuse. {\em Astrophys. J.} {\bf 2020}, {\em 899}, 68--79. [\href{http://dx.doi.org/10.3847/1538-4357/aba516}{CrossRef}]

\bibitem[Humphreys, et al.(2021)]{RMH2021}
Humphreys, R.M.; Davidson, K.; Richards, A.M.S.; Ziurys, L.M.; Jones, T.J.; Ishibashi, K. The Mass-loss History of the Red Hypergiant VY CMa. {\em Astron. J.} {\bf 2021}, {\em 161}, 98--108. [\href{http://dx.doi.org/10.3847/1538-3881/abd316}{CrossRef}]

\bibitem[Humphreys, et al.(2024)]{RMH2024}
Humphreys, R.M.; Richards, A.M.S.; Davidson, K.; Singh, A.P.; Decin, L.; Ziurys, L.M. The Hidden Clumps in VY CMa Uncovered by ALMA. {\em Astron. J.} {\bf 2024}, {\em 167}, 94--106. [\href{http://dx.doi.org/10.3847/1538-3881/ad1dd7}{CrossRef}]

\bibitem[Humphreys \& Jones(2022)]{HJ}
Humphreys, R.M.; Jones, T.J. Episodic Gaseous Outflows and Mass Loss from Red Supergiants.  {\em Astron. J.} {\bf 2022}, {\em 163}, 103--109. [\href{http://dx.doi.org/10.3847/1538-3881/ac46f}{CrossRef}]

\bibitem[Vlemmings et al.(2002)]{Vlemmings}
Vlemmings, W.H.T.; Diamond, P.J.; van Langevelde, H.J. Circular Polarization of Water Masers in the Circumstellar Envelopes of Late Type Stars. {\em Astron. Astrophys.} {\bf 2002}, {\em 394}, 589--602. [\href{http://dx.doi.org/10.1051/0004-6361:20021166}{CrossRef}]

\bibitem[Zwicky(1965)]{Zwicky}
Zwicky, F. Supernovae. In {\it Stellar Structure--Stars and Stellar Systems  Vol. VIII};  Aller, L.H., McLaughlin, D.B., Eds.; University of Chicgo Press: Chicago, IL, USA, {1965}; p. 367.

\bibitem[Minkowski(1941)]{Minkowski}
Minkowski, R. Spectra of Supernovae. {\em Publ. Astron. Soc. Pac.} {\bf 1941}, {\em 53}, 224--225. [\href{http://dx.doi.org/10.1086/125315}{CrossRef}]


\bibitem[Barbon et al.(1965)]{Barbon}
Barbon, R.; Dallaporta, N.; Perinott, M.; Sussi, M.G.  On the lower mass limit for implosion type Supernovae. {\em Mem. Soc. Astronom. Ital.} {\bf 1965}, {\em 36}, 127--150.

\bibitem[Barbaro et al.(1969)]{Barbaro}
Barbaro, G.; Dallaporta, N.; Summa, G. On the Presumed presupernova Stage for type II Supernovae. {\em Comm. Konkoly Obs.} {\bf 1969}, {\em 65}, 41--50.


\bibitem[Barbon et al.(1979)]{Barbon79}
Barbon, R.; Ciatti, F.; Rosino, L. Photometric Properties of Type II Supernovae. {\em Astron. Astrophys.} {\bf 1979}, {\em 72}, 287--292.

\bibitem[Maeder(1981)]{Maeder}
Maeder, A. The most massive stars evolving to red supergiants---Evolution with mass loss, WR stars as post-red supergiants and pre-supernovae. {\em Astron. Astrophys.} {\bf 1981}, {\em 99}, 97--107.


\bibitem[Smartt et al.(2009)]{Smartt2009}
Smartt, S.J.; Eldridge, J.J.; Crockett, R.M.; Maund, J.R.  The death of massive
stars---I. Observational constraints on the progenitors of Type II-P supernovae.
{\em Mon. Not. R. Astron. Soc.} {\bf 2009}, {\em 395}, 1409--1437. [\href{http://dx.doi.org/10.1111/j.1365-2966.2009.14}{CrossRef}]

\bibitem[Smartt(2015)]{Smartt2015}
Smartt, S.J. Observational Constraints on the Progenitors of Core-Collapse Supe rnovae: The Case for Missing High-Mass Stars. {\em Publ. Astron. Soc. Aust.} {\bf 2015}, {\em 32}, e002. [\href{http://dx.doi.org/10.1017/pasa.2015.1}{CrossRef}]

\bibitem[Beasor \& Davies(2016)]{Beasor1}
Beasor, E.M.; Davies, B. The Evolution of Red Supergiants to Supernova in NGC 2100. {\em Mon. Not. R. Astron. Soc.} {\bf 2016}, {\em 463}, 1269--1283. [\href{http://dx.doi.org/10.1093/mnras/stw2054}{CrossRef}]

\bibitem[Davies \& Beasor(2018)]{Beasor2}
Davies, B.; Beasor, E.M. The Initial Masses of the Red Supergiant Progenitors to Type II Supernovae. {\em Mon. Not. R. Astron. Soc.} {\bf 2018}, {\em 474}, 2116--2128. [\href{http://dx.doi.org/10.1093/mnras/stx2734}{CrossRef}]

\bibitem[Walmswell \& Eldridge(2012)]{Walmswell}
Walmswell, J.J.; Eldridge, J.J. Circumstellar Dust as a Solution to the Red Supergiant Supernova Progenitor Problem. {\em Mon. Not. R. Astron. Soc.} {\bf 2012}, {\em 419}, 2054--2062. [\href{http://dx.doi.org/10.1111/j.1365-2966.2011.19860.x}{CrossRef}]

\bibitem[Kilpatrick \& Foley(2018)]{Kilpatrick}
Kilpatrick, C.D.; Foley, R.J. The Dusty Progenitor Star of the Type II Supernova 2017eaw. {\em Mon. Not. R. Astron. Soc.} {\bf 2018}, {\em 481}, 2536--2547. [\href{http://dx.doi.org/10.1093/mnras/sty2435}{CrossRef}]

\bibitem[Gerke et al.(2015)]{Gerke}
Gerke, J.R.; Kochanek, C.S.; Stanek, K.Z. The Search for Failed Supernovae with the Large Binocular Telescope: First Candidates. {\em Mon. Not. R. Astron. Soc.} {\bf 2015}, {\em 450}, 3289--3305. [\href{http://dx.doi.org/10.1093/mnras/stv776}{CrossRef}]

\bibitem[Adams et al.(2018)]{Adams}
Adams, S.M.; Kochanek, C.S.; Gerke, J.R.; Stanek, K.Z.; Dai, X. The Search for Failed Supernovae with the Large Binocular Telescope: Confirmation of a Disappearing Star. {\em Mon. Not. R. Astron. Soc.} {\bf 2017}, {\em 468}, 4968--4981. [\href{http://dx.doi.org/10.1093/mnras/stx816}{CrossRef}]



\bibitem[Humphreys(1019)]{RH-BH1}
Humphreys, R.M. Comments on the Progenitor of NGC 6946-BH1. {\em Res. Notes AAS} {\bf 2019}, {\em 3}, 164--167. [\href{http://dx.doi.org/10.3847/2515-5172/ab5191}{CrossRef}]

\bibitem[De et al.(2024)]{De}
De, K.; MacLeod, M.; Jencson, J.E.; Lovegrove, E.; Antoni, A.; Kara, E.; Kasliwal, M.M.; Lau, R.M.; Loeb, A.; Masterson, M.; et al. The Disappearance of a Massive Star Marking the Birth of a Black Hole in M31.  \emph{arXiv} \textbf{2024}, arXiv:2410.14778.

\bibitem[Humphreys, et al.(2006)]{RMH06}
Humphreys, R.M.; Jones, T.J.; Polomski, E.; Koppelman, M.; Helton, A.; McQuinn, K.; Gehrz, R.D.; Woodward, C.E.; Wagner, R.M.; Gordon, K.; et al. M33's Variable A: A Hypergiant Star More Than 35 YEARS in Eruption. {\em Astron. J.} {\bf 2006}, {\em 131}, 2105--2113. [\href{http://dx.doi.org/10.1086/500811}{CrossRef}]

\bibitem[Tiffany et al.(2010)]{Tiffany}
Tiffany, C.; Humphreys, R.M.; Jones, T.J.; Davidson, K. The Morphology of IRC+10420's Circumstellar Ejecta. {\em Astron. J.} {\bf 2010}, {\em 140}, 339--349. [\href{http://dx.doi.org/10.1088/0004-6256/140/2/339}{CrossRef}]

\bibitem[Shenoy, et al.(2016)]{Shenoy}
Shenoy, D.; Humphreys, R.M.; Jones, T.J.; Marengo, M.; Gehrz, R.D.; Helton, L.A.; Hoffmann, W.F.; Skemer, A.J.; Hinz, P.M. Searching for Cool Dust in the Mid-to-far Infrared: The Mass-loss Histories of the Hypergiants $\mu$ Cep, VY CMa, IRC+10420, and $\rho$ Cas. {\em Astron. J.} {\bf 2016}, {\em 151}, 51--65. [\href{http://dx.doi.org/10.3847/0004-6256/151/3/51}{CrossRef}]

\bibitem[Humphreys, et al.(1997)]{RMH1997}
Humphreys, R.M.; Smith, N.; Davidson, K.; Jones, T.J.; Gehrz, R.T.; Mason, C.G.; Hayward, T.L.; Houck, J.R.; Krautter, J. HST and Infrared Images of the Circumstellar Environment of the Cool Hypergiant IRC + 10420. {\em Astron. J.} {\bf 1997}, {\em 114}, 2778--2788. [\href{http://dx.doi.org/10.1086/118686}{CrossRef}]

\bibitem[Smith, et al.(2001)]{Smith}
Smith, N.; Humphreys, R.M.; Davidson, K.; Gehrz, R.D.; Schuster, M.T.; Krautter, J. The Asymmetric Nebula Surrounding the Extreme Red Supergiant VY Canis Majoris. {\em Astron. J.} {\bf 2001}, {\em 121}, 1111--1125. [\href{http://dx.doi.org/10.1086/318748}{CrossRef}]

\bibitem[Schuster, et al.(2006)]{Schuster}
Schuster, M.T.; Humphreys, R.M.; Marengo, M. The Circumstellar Environments of NML Cygni and the Cool Hypergiants. {\em Astron. J.} {\bf 2006}, {\em 131}, 603--611. [\href{http://dx.doi.org/10.1086/498395}{CrossRef}]


\bibitem[Maund, et al.(2011)]{Maund}
Maund, J.R.; Fraser, M.; Ergon, M.; Pastorello, A.; Smartt, S.J.; Sollerman, J.; Benetti, S.; Botticella, M.-T.; Bufano, F.; Danziger, I.J.; et al. The Yellow Supergiant Progenitor of the Type II Supernova 2011dh in M51.  {\em Astrophys. J.} {\bf 2011}, {\em 739}, 37L--42L. [\href{http://dx.doi.org/10.1088/2041-8205/739/2/l37}{CrossRef}]

\bibitem[Van Dyk, et al. (2011)]{vandyk11}
Van Dyk, S.D.; Li, W.; Cenko, S.B.; Kasliwal, M.M.; Horesh, A.; Ofek, E.O.; Kraus, A.L.; Silverman, J.M.; Arcavi, I.; Filippenko, A.V.; et al. The Progenitor of Supernova 2011dh/PTF11eon in Messier 51. {\em Astrophys. J.} {\bf 2011}, {\em 741}, L28--L33. [\href{http://dx.doi.org/10.1088/2041-8205/741/2/l28}{CrossRef}]

\bibitem[Van Dyk, et al.(2013)]{vandyk13}
Van Dyk, S.D.; Zheng, W.; Clubb, K.I.; Filippenko, A.V.; Cenko, S.B.; Smith, N.; Fox, O.D.; Kelly, P.L.; Shivvers, I.; Ganeshalingam, M. The Progenitor of Supernova 2011dh has Vanished. {\em Astrophys. J.} {\bf 2013}, {\em 772}, L32--L37. [\href{http://dx.doi.org/10.1088/2041-8205/772/2/L32}{CrossRef}]

\end{thebibliography}
\end{document}